\documentclass[9pt,twocolumn,twoside]{osajnlarx}

\journal{josab} 
\setboolean{shortarticle}{true} 

\title{Effects of loss on the phase sensitivity with parity detection in an SU(1,1) interferometer}

\author[1,2]{Dong Li}
\author[3,5,*]{Chun-Hua Yuan}
\author[1,2]{Yao Yao}
\author[1,2]{Wei Jiang}
\author[1,2,$\dagger$]{Mo Li}
\author[4,5]{Weiping Zhang}

\affil[1]{Microsystems and Terahertz Research Center, China Academy of Engineering Physics,
Chengdu Sichuan 610200, P. R. China}
\affil[2]{Institute of Electronic Engineering, China Academy of Engineering Physics, Mianyang Sichuan 621999, P. R. China}
\affil[3]{Quantum Institute for Light and Atoms, Department of Physics, East China Normal University, Shanghai 200062, P. R. China}
\affil[4]{Department
of Physics, Shanghai Jiao Tong University, and Tsung-Dao Lee Institute, Shanghai 200240, P. R. China}
\affil[5]{Collaborative Innovation Center of Extreme Optics, Shanxi University, Taiyuan, Shanxi 030006, P. R. China}

\affil[*]{Corresponding author: chyuan@phy.ecnu.edu.cn}
\affil[$\dagger$]{Corresponding author: limo@mtrc.ac.cn}

\ociscodes{(270.0270) Quantum optics; (120.5050) Phase measurement; (120.3940) Metrology; (120.3180) Interferometry.}

\begin{abstract}
We theoretically study the effects of loss on the phase sensitivity of an SU(1,1) interferometer with parity detection with various input states. We show that although the sensitivity of phase estimation decreases in the presence of loss, it can still beat the shot-noise limit with small loss. To examine the performance of parity detection, the comparison is performed among homodyne detection, intensity detection, and parity detection. Compared with homodyne detection and intensity detection, parity detection has a slight better optimal phase sensitivity in the absence of loss, but has a worse optimal phase sensitivity with a significant amount of loss with one-coherent state or coherent $\otimes$ squeezed state input. 
\end{abstract}

\setboolean{displaycopyright}{true}

\begin{document}

\maketitle

\section{Introduction}

Quantum optical interferometer, a primary tool for various precision
measurements, has long been proposed to achieve higher sensitivity than what is possible classically~\cite%
{helstrom1976quantum,holevo2011probabilistic,caves81,xiao1987precision,caves94,demkowicz2012elusive,lee2002quantum,giovannetti2004quantum,Gao:10,dowling2008quantum,Berrada:17,szigeti2017,anders2017}%
. Recently, physicists with the advanced Laser Interferometer
Gravitational-Wave Observatory (LIGO) observed the gravitational waves~\cite%
{abbott2016gw150914} owing to the development of the advanced optical interferometric
measurement technology. The optical interferometer works based on mapping
the quantity of interest onto the phase variance of a system and estimating
the latter, for example, the relative phase between the two modes or
 "arms" of an interferometer~\cite{kaushik13}. However, the
phase sensitivity of a classical measurement scheme is limited by the
shot-noise limit (SNL), $1/\sqrt{\bar{N}}$, where $\bar{N}$ is the mean
total photon number inside the interferometer.

In the quantum optical metrology, one goal is to achieve a sensitivity of phase estimation below the SNL. For this purpose, in 1986 Yurke \emph{et al.}~\cite%
{yurke86} proposed a theoretical scheme of the SU(1,1) interferometer. In this type of interferometer, the splitter and recombination of the beams are done through nonlinear interactions while conventional SU(2) interferometers use linear beam splitters. They showed that such a kind of interferometer provides the potential of achieving improved sensitivity of phase estimation. This is because of reduction of the noise and amplification of signal achieved by the nonlinear interactions.

Recently, the experimental realization of such a
nonlinear interferometer was reported by Jing \emph{et al}.~\cite{jing2011}
in which the nonlinear beam splitters are realized by using optical parametric amplifiers (OPAs) and the maximum output intensity can be much higher than the input due to the parametric amplification. In 2014, an improvement of 4.1 dB in
signal-to-noise ratio was observed by Hudelist \emph{et al}.~\cite%
{hudelist2014quantum} compared with the SU(2) interferometer under the same
operation condition. In 2017, Anderson \emph{et al}.~\cite{Anderson:17} observed that the "truncated SU(1,1) interferometer" can surpass the SNL by $4$ dB even with $\approx 30\%$ loss. 


The discussion above focuses only on the SU(1,1) interferometers realized by using OPAs as beam splitters experimentally which can be characterized by all-optical ones. In contrast, another kinds of SU(1,1) interferometers have been also realized experimentally. For example, the atom-light hybrid SU(1,1)
interferometer~\cite{bing15,ma20151,chen2016effects} has been reported which
used the interface between the atomic pseudospin wave and light.
Besides, the all-atomic SU(1,1) interferometer~\cite{gross2010nonlinear, peise2015interaction,
gabbrielli2015spin,linnemann2016,linnemann2017,peise2015,kruse2016} was also studied. Gabbrielli
\emph{et al}.~\cite{gabbrielli2015spin} presented a nonlinear three-mode
SU(1,1) atomic interferometer realized with ultracold atoms. Furthermore,
Barzanjeh \emph{et al}.~\cite{barzanjeh2014dispersive} proposed to achieve the
SU(1,1)-type interferometer by using the circuit quantum electrodynamics system. More recently, Chekhova \emph{et al}.~\cite{Chekhova:16} presented a detail review of the progress on the field of SU(1,1) interferometer for its application in precision metrology.

However, in realistic systems the interaction with the environment is inevitable. Since
quantum procedures are susceptible to noise, the analysis of phase estimation in noisy
or dissipative environment is required~\cite%
{kebei12,lee2009optimization,loss2010,gong2017}. From this point of view, the effects of loss on the phase sensitivity of the SU(1,1)
interferometers were investigated by Marino \emph{et al.}~\cite{marino12} in 2012 where the measurement scheme considered was intensity detection (ID). They showed that the phase sensitivity can still surpass the SNL possibly with the loss being smaller than $50\%$, even though the photon losses
degrade the sensitivity of phase estimation. In 2014, some of the authors~\cite{li2014phase}
also studied the effects of loss on the performance in an SU(1,1) interferometer via homodyne detection (HD). They presented that the photon losses would reduce the sensitivity of phase estimation where the effects of the loss between the two OPAs on the phase sensitivity is greater than the loss after the second OPA when $\cosh(2g) > 1$ with $g$ being the OPA strength. 

It is worthy noting that the HD measures the quadrature of the field \cite{walls2007quantum} which is different from the intensity detection monitoring the mean total photon number of the field with its corresponding photon number operator $\hat{N}_c \equiv \hat{c}^{\dagger} \hat{c}$ ($\hat{c}$ and $\hat{c}^{\dagger}$ are annihalation and creation operators of mode $c$, respectively). In general, it is defined as $\hat{X}_c(\theta) = (\hat{c}e^{-i\theta} + \hat{c}^{\dagger}e^{i\theta})/\sqrt{2}$ for the quadrature operator of mode $c$ where $\theta$ determines the quadrature phase. When $\theta = 0$, it is reduced to $\hat{X}_{c}(0) = (\hat{c} + \hat{c}^{\dagger}) / \sqrt{2}$ which is usually called as dimensionless position operator while $\hat{X}_{c}(\pi/2) = (\hat{c} - \hat{c}^{\dagger})/(i\sqrt{2})$ is called as dimensionless momentum operator when $\theta = \pi/2$ \cite{walls2007quantum}. Specifically, the homodyne detection is taken along the $\hat{X}_{c}(0)$ quadrature in the scheme considered here. Moreover, for a balanced homodyne detection scheme, one would impinge one of the outgoing light outputs onto a 50:50 linear beam splitter, along with a coherent state of the same frequency as the input coherent state and perform intensity difference between the two outputs of this beam splitters which is a mature technique for quantum optical measurement experiments nowadays \cite{bachor2004}. 

As described above, the effects of loss with the HD and the ID have been analyzed in an SU(1,1) interferometer. However, the analysis of the effects of loss with the parity detection (PD) on an SU(1,1) interferometer is still missing. The parity detection serves as an optimal detection strategy when the given states are subject to the phase fluctuations~\cite{Bardhan:2013}. The effects of loss on the
performance of an SU(1,1) interferometer via the PD also merit investigation. In 2016, some of the authors~\cite{li2016phase} analyzed the performance of the PD in an SU(1,1) interferometer in the absence of loss. In this paper, we will investigate the effects of loss on the phase sensitivity
in an SU(1,1) interferometer with the PD. In the presence of small
loss, the phase sensitivity will be reduced but still surpass the SNL. Moreover, we also compare the phase sensitivities among the HD, the ID, and the PD in a lossy SU(1,1) interferometer. Furthermore, the phase sensitivity with parity detection is compared with the quantum Cram\'er-Rao bound (QCRB) \cite{helstrom1976quantum,caves94}, which gives the ultimate limit for a set of probabilities that originated from measurements on a quantum system. The QCRB can be obtained by the maximum likelihood estimator and presents a measurement-independent phase sensitivity $\Delta \phi_{\text{QCRB}}$.

\begin{figure}[tbp]
\centerline{\includegraphics[scale=0.95,angle=0]{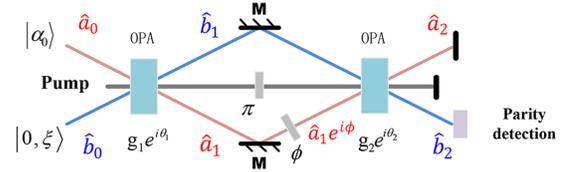}}
\caption{An SU(1,1) interferometer with a two-mode input in
which two OPAs take the place of two beam splitters in the traditional
Mach-Zehnder interferometer. $g_{1}$ ($g_{2}$) and $\protect\theta _{1}$ ($%
\protect\theta _{2}$) describe the strength and phase shift in the OPA
process $1$ ($2$), respectively. $\hat{a}_{i}$ and $\hat{b}_{i}$ ($i=0,1,2$) mean the annihilation operators of modes $a$ and $b$, respectively. The pump field between the two OPAs
has a $\pi $ phase difference compared with the pump field before the first OPA. The output
of mode $b_{2}$ is measured by parity detection. {\bf{M}}: mirrors, $\phi $: phase shift.}
\label{fig1}
\end{figure}

This manuscript is organized as follows: section 2 presents the phase sensitivity in ideal case and the loss effects on the phase sensitivity with the PD which is followed by a comparison among different detections in section 3. Last we conclude with a summary.

\section{Phase sensitivity in an SU(1,1) interferometer}
\subsection{Ideal case}
\label{sec:idealcase}
The PD was first proposed
by Bollinger \emph{et al}.~\cite{Bollinger96} to study spectroscopy of trapped ions by counting their number in 1996. Later Gerry~\cite{gerry00}
introduced the PD into optical interferometers. The PD on an SU(2) interferometer has been proven to be an efficient
measurement method for a wide range of input states~\cite%
{gerry03,gerry05,gerry05b,gerry2007parity} where the PD works as well, or nearly as well, as state-specific detection schemes~\cite%
{gerry2010parity,chiruvelli2011parity}. Mathematically, parity here means simply the evenness or
oddness of the photon number in an output mode. Its corresponding parity operator
is described by $\hat{\Pi}=(-1)^{\hat{N}}$ with $\hat{N}$ being the single-mode photon number operator. In experiments, the PD can be achieved by using homodyne
techniques \cite{plick2010parity} for high power beam, or counting the photon number with a
photon-number resolving detector directly for low optical power \cite{cohen2010}.

In our case, the parity operator of the output mode $b$ is naturally defined as $\hat{\Pi}%
_{b}\equiv (-1)^{\hat{b}_{2}^{\dag }\hat{b}_{2}}$ where $\hat{b}_2 $ and $\hat{b}_2^{\dagger} $ are annihilation and creation operators of output mode $b$, respectively. According to the linear error propagation formula, the sensitivity of phase
estimation with the PD is given by%
\begin{equation}
\Delta \phi =\frac{\langle \Delta \hat{\Pi}_{b}\rangle }{%
\left\vert \frac{\partial \langle \hat{\Pi}_{b}\rangle }{\partial \phi }%
\right\vert },  \label{esti}
\end{equation}%
where $\langle \Delta \hat{\Pi%
}_{b}\rangle \equiv (\langle \hat{\Pi}_{b}^{2}\rangle -\langle \hat{\Pi}%
_{b}\rangle ^{2})^{1/2}=(1-\langle \hat{\Pi}_{b}\rangle ^{2})^{1/2}.$

We consider a coherent and
squeezed vacuum state as input here. Transport of input fields through an SU(1,1) interferometer is described in Appendix~\ref{model}. Based on this model, $\langle \hat{\Pi}_{b}\rangle $ is worked out as a series of rather complex
and un-illuminating expressions as shown in Appendix~\ref{sec:signal}. Then
the phase sensitivity in an SU(1,1) interferometer with a coherent and
squeezed vacuum input state with the PD is found to be minimal at $%
\phi =0$ and is given by%
\begin{equation}
\Delta \phi =\frac{1}{G_{\mathrm{OPA}}}\frac{1}{\left\{
N_{\alpha }[\sinh (2r)\cos (2\theta _{\alpha })+\cosh (2r)]+N_{s}+1\right\}
^{1/2}},
\end{equation}%
where $N_{\alpha }=|\alpha _{0}|^{2}$ is the mean total photon number of
input coherent state with $\alpha_0 = |\alpha_0| e^{i \theta_{\alpha}}$ being the amplitude of
input coherent state, $N_{s}=\sinh ^{2}r$ is the intensity of the squeezed
vacuum light with its squeezing parameter $r$, and $G_{\mathrm{OPA}}=\sqrt{N_{\mathrm{OPA}}(N_{\mathrm{OPA}%
}+2)}=\sinh (2g)$ with $N_{\mathrm{OPA}}=2\sinh ^{2}g$ being the spontaneous photon
number emitted from the first OPA with the parametric strength of $g$ (we use $g_1 = g_2 = g$).
When $\theta _{\alpha }=0,$ the optimal phase sensitivity is obtained and is found to be%
\begin{equation}
\Delta \phi =\frac{1}{G_{\mathrm{OPA}}}\frac{1}{[(N_{\alpha
}e^{2r}+N_{s}+1)]^{1/2}},  \label{P}
\end{equation}%
where the factor $e^{2r}$ results from the input squeezed vacuum beam. If vacuum input ($N_{s}=0$ and $N_{\alpha }=0$) the
phase sensitivity with the PD is then simplified to $\Delta \phi =1/G_{\mathrm{OPA}}$ which is the same as the result of the ID~\cite{yurke86}. Thus, the PD has the same optimal scaling of phase sensitivity as the ID with vacuum input. 

For the sake of clarity, the corresponding Heisenberg limit (HL) \cite{marino12} is presented in Appendix~\ref{sec:hl} and is given by
\begin{equation}
\Delta \phi _{\mathrm{HL}}=\frac{1}{N_{\mathrm{Tot}}},  \label{HL}
\end{equation}%
where $N_{\mathrm{Tot}}=(N_{\mathrm{OPA}}+1)N_{\mathrm{in}}+N_{\mathrm{OPA}}$ is the total mean photon number inside the nonlinear interferometer with $N_{\mathrm{in}}=N_{\alpha }+N_{s}$ being the total mean input photon number~\cite{li2014phase}. Meanwhile, the corresponding shot-noise limit (SNL) is given by $\Delta \phi_{\text{SNL}} = 1/ \sqrt{N_{\text{Tot}}}$ \cite{marino12}. According to Ref.~\cite{li2016phase}, it is easy to find another quantum limit, the QCRB of an SU(1,1) interferometer, which is shown in Appendix~\ref{sec:qcrb}.

\subsection{Effects of loss}

The decoherence process is
inevitable by the imperfections in interferometric setups. As discussed in reference~\cite{demkowicz2015chapter}, there are
various kinds of decoherence processes, such as, the effects of phase
diffusion, the impact of imperfect visibility and photonic losses. For simplicity, here we only consider the effects of photonic losses. It is well known
that loss has a significant effect on the phase sensitivity~\cite%
{ou2012enhancement,gilbert2008use,dorner2009optimal}. In the presence of
dissipation, the performance of phase estimation is degraded due to the
loss of photons. For this reason, any practical implementation of the
measurement scheme must pay special attention to the loss effects carefully.

\begin{figure}[tb!]
\centerline{\includegraphics[scale=0.45,angle=0]{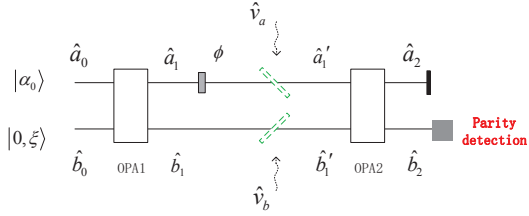}}
\caption{A lossy interferometer model, the losses in the
interferometer are modeled by adding fictitious beam splitters. $\hat{v}_a $ and $\hat{v}_b $ denote the vacuum modes $v_a$ and $v_b$, respectively. }
\label{fig2}
\end{figure}

In this section, we investigate the effects of loss on the phase sensitivity
in an SU(1,1) interferometer with the PD. Theoretically, photon
loss is typically modeled by a beam splitter that routes photons out of the
interferometer as shown in Fig.~\ref{fig2} \cite{lee2009optimization}. $%
L_{1}$ and $L_{2}$ describe the loss on the up and bottom arms,
respectively. 

With this scheme, the pure-state input is now described as a Wigner function
for the four modes including two vacuum modes. Due to two external vacuum modes introduced, the initial
Wigner function for the four modes is entirely described as%
\begin{align}
W_{\mathrm{in}}(\alpha _{i}^{\prime };\beta _{i}^{\prime
};v_{a0};v_{b0})=&W_{|\alpha _{0}\rangle }(\alpha _{i},\alpha _{0})W_{|0,\xi
\rangle }(\beta _{i},r)\nonumber\\ 
& \times W_{|0\rangle _{a0}}(v_{a0})W_{|0\rangle
_{b0}}(v_{b0}),
\label{inwloss}
\end{align}%
where $W_{|0\rangle _{a0}}(v_{a0})$ and $W_{|0\rangle _{b0}}(v_{b0})$ are
Wigner functions of vacuum in $v_{a}$ and $v_{b}$ modes, respectively, and are given by \cite{walls2007quantum}
\begin{align}
W_{|0\rangle _{a0}}(v_{a0}) = \frac{2}{\pi} e^{-2|v_{a0}|^2},\nonumber\\
W_{|0\rangle _{b0}}(v_{b0}) = \frac{2}{\pi} e^{-2|v_{b0}|^2},
\label{inwvac}
\end{align}
where the complex numbers $v_{a0}$ and $v_{b0}$ are introduced and distributed over the Wigner phase space.

Similar to the ideal case as shown in Appendix~\ref{model}, after passing through the nonlinear interferometer, the corresponding output Wigner function can be then written as%
\begin{align}
W_{\mathrm{out}}(\alpha _{f}^{\prime };\beta _{f}^{\prime };v_{af};v_{bf})
=&W_{\mathrm{in}}[\alpha _{i}^{\prime }(\alpha _{f}^{\prime };\beta
_{f}^{\prime };v_{af};v_{bf});\beta _{i}^{\prime }(\alpha _{f}^{\prime
};\beta _{f}^{\prime };v_{af};v_{bf});  \nonumber \\
&v_{a0}(\alpha _{f}^{\prime };\beta _{f}^{\prime
};v_{af};v_{bf});v_{b0}(\alpha _{f}^{\prime };\beta _{f}^{\prime
};v_{af};v_{bf})].  \label{outwloss}
\end{align}%
Crucially, the four-mode-input-output relation of the SU(1,1) interferometer is given by%
\begin{equation}
\left(
\begin{array}{c}
\alpha _{i} \\
\beta _{i}^{\ast } \\
v_{a0} \\
v_{b0}^{\ast }%
\end{array}%
\right) =T^{\prime -1}\left(
\begin{array}{c}
\alpha _{f} \\
\beta _{f}^{\ast } \\
v_{af} \\
v_{bf}^{\ast }%
\end{array}%
\right) ,  \label{inputoutloss}
\end{equation}%
where the total transformation matrix is found to be
\begin{equation}
T^{\prime }=T_{\mathrm{OPA1}%
}^{\prime }T_{\phi }^{\prime }T_{\mathrm{loss}}^{\prime }T%
_{\mathrm{OPA2}}^{\prime },\nonumber
\end{equation} 
with%
\begin{align}
T_{\mathrm{OPA1}}^{\prime } =&\left(
\begin{array}{cccc}
u_{1} & v_{1} & 0 & 0 \\
v_{1}^{\ast } & u_{1} & 0 & 0 \\
0 & 0 & 1 & 0 \\
0 & 0 & 0 & 1%
\end{array}%
\right),\nonumber \\
T_{\phi }^{\prime }=&\left(
\begin{array}{cccc}
e^{i\phi } & 0 & 0 & 0 \\
0 & 1 & 0 & 0 \\
0 & 0 & 1 & 0 \\
0 & 0 & 0 & 1%
\end{array}%
\right), \nonumber \\
T_{\mathrm{loss}}^{\prime } =&\left(
\begin{array}{cccc}
\sqrt{1-L_{1}} & 0 & \sqrt{L_{1}} & 0 \\
0 & \sqrt{1-L_{2}} & 0 & \sqrt{L_{2}} \\
\sqrt{L_{1}} & 0 & \sqrt{1-L_{1}} & 0 \\
0 & \sqrt{L_{2}} & 0 & \sqrt{1-L_{2}}%
\end{array}%
\right), \nonumber \\
T_{\mathrm{OPA2}}^{\prime } =&\left(
\begin{array}{cccc}
u_{2} & v_{2} & 0 & 0 \\
v_{2}^{\ast } & u_{2} & 0 & 0 \\
0 & 0 & 1 & 0 \\
0 & 0 & 0 & 1%
\end{array}%
\right),
\end{align}
where $u_{k}=\cosh g_{k}, v_{k}=e^{i\theta _{k}}\sinh g_{k}$, and $v_{k}^{\ast}$ is the conjugate of $v_{k}$ $ (k = 1,2)$, $\theta _{1}$ $(\theta _{2})$ and $g_{1}$ $(g_{2})$ are the phase shift and parametric strength in the OPA
process 1 (2), respectively. $L_1$ and $L_2$ denote the photon loss on the modes $a_1$ and $b_1$ inside the interferometer, respectively.
\begin{figure}[tb!]
\includegraphics[scale=0.5]{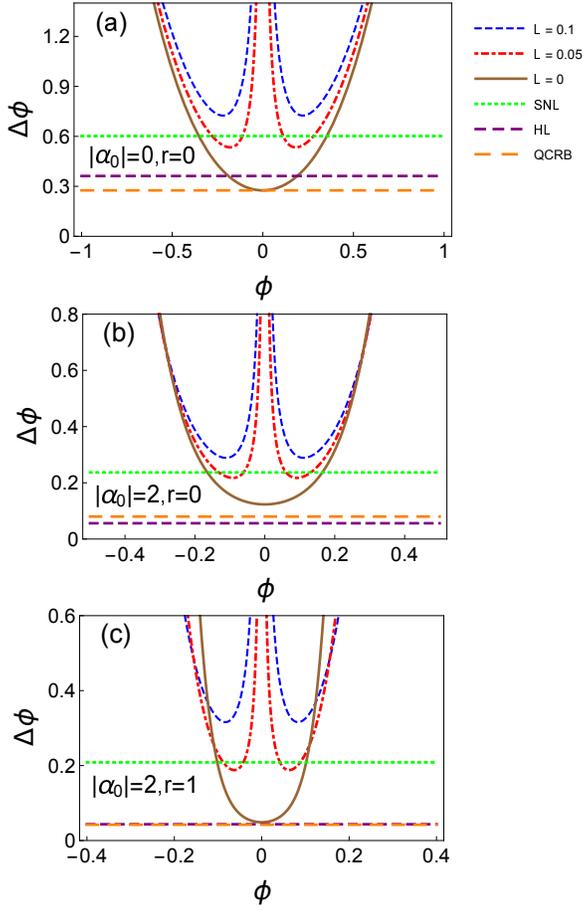}
\caption{Sensitivity of phase estimation with the PD versus $\protect\phi$ with various inputs of (a) vacuum state ($|%
\protect\alpha_0| = 0$ and $r = 0$); (b) only one coherent state ($|\protect%
\alpha_0| = 2$ and $r = 0$); (c) coherent and squeezed vacuum state ($|%
\protect\alpha_0| = 2$ and $r = 1$), and with different cases of loss: the
dashed-blue line is for $L = 0.1$, the dash-dotted-red line for $L = 0.05$,
the solid-brown line for the ideal case, and the dotted-green,
dashed-purple, and dashed-orange lines for the SNL, the HL, and the QCRB, respectively. Parameters used are: $L_1 =
L_2 = L$ and $g = 1$.}
\label{phi}
\end{figure}

By combining Eqs. (\ref{inwloss}), (\ref{inwvac}), (\ref{outwloss}), (\ref{inputoutloss}), and (\ref{signal}) shown in Appendix \ref{model}, the signal of the PD on the output of mode $b_2$ is given by%
\begin{equation}
\langle \hat{\Pi}_{b}^{\mathrm{loss}}\rangle =\frac{\pi }{2} \int W_{\mathrm{%
out}}(\alpha _{f}^{\prime };0;v_{af};v_{bf})d^{2}\alpha _{f}^{\prime
}d^{2}v_{af}d^{2}v_{bf}.  \label{wloss}
\end{equation}%
The result of $\langle \hat{\Pi}_{b}^{\mathrm{loss}}\rangle $ is a series of
rather large equations, which is reported in Appendix~\ref{sec:signalloss}. Then
similar to Eq.~(\ref{esti}) the phase sensitivity $\Delta \phi _{%
\mathrm{L}}$ in the presence of loss is worked out by%
\begin{equation}
\Delta \phi _{\mathrm{L}}=\frac{\langle \Delta \hat{\Pi}_{b}^{\mathrm{loss%
}}\rangle }{\left\vert \frac{\partial \langle \hat{\Pi}_{b}^{\mathrm{loss}%
}\rangle }{\partial \phi }\right\vert },  \label{sensiloss}
\end{equation}%
where the subscript L denotes the loss and the phase sensitivity $\Delta \phi _{\mathrm{L}}$ is also a series of complex expressions and is not reported here.

So far, we have obtained the performance of a nonlinear
interferometer via the PD in the presence of loss. Specifically, if vacuum inputs ($\alpha
_{0}=0,r=0$), $\Delta \phi _{\mathrm{L}}$ is simplified to%
\begin{align}
\Delta \phi _{\mathrm{L}} =&\{1-[(1-L)\sinh ^{2}(2g)\cos \phi -(1-L)\cosh^2 (2g)\nonumber \\
&-L\cosh
(2g)]^{-2}\}^{1/2}
 \frac{\csc \phi }{16(1-L)}\{4\coth
(2g)[L  \nonumber \\
&+(1-L)\cosh (2g)]   -4(1-L)\sinh (2g)\cos \phi \}^{2},
\label{senvacloss}
\end{align}%
where we have set $L_{1}=L_{2}=L$. In the absence of loss ($L=0$) and at the
optimal phase point ($\phi _{\mathrm{opt}}=0$) the optimal phase sensitivity
in Eq.~(\ref{senvacloss}) can be reduced to%
\begin{equation}
\Delta \phi _{\mathrm{L}}=\frac{1}{G_{\mathrm{OPA}}} = \frac{1}{\sqrt{N_{\text{OPA}} (N_{\text{OPA}} + 2)}},
\label{senvac}
\end{equation}%
which agrees with the result of ideal case. 

It shows the phase sensitivity with the PD as a
function of $\phi $ for various cases of loss ($L=0.1,L=0.05$, and $L=0$)
under the condition of vacuum state input [one-coherent-state input and coherent $\otimes$ squeezed-vacuum state input] as depicted in Fig.~\ref{phi}(a) [\ref{phi}(b) and \ref{phi}(c)]. One can easily see that the
optimal phase point is at $\phi _{\mathrm{opt}}=0$ in the absence of loss in Figs. \ref{phi}(a), \ref{phi}(b), and \ref{phi}(c). However, considering the
effects of loss, $\phi _{\mathrm{opt}}$ becomes apart from $0$. Additionally,
the optimal phase point tends to be far away from $0$ with loss increasing. This is due to decorrelation point ($\phi = 0$) playing a significant role on precision phase estimation~\cite{chen2016effects}. At nearby the decorrelation point, the detection noise is amplified a little and the optimal phase sensitivity is then obtained. Without loss, $\phi = 0$ is the decorrelation point leading that the outputs have the same correlation as inputs. However, $\phi = 0$ is not the decorrelation point when considering the effects of photon loss~\cite{chen2016effects} which decreases the precision of phase estimation.

In Fig. \ref{phi}(a), in the absence of loss, it reaches below the HL ($\Delta \phi_{\text{HL}} = 1/N_{\text{OPA}}$) for the optimal sensitivity of phase estimation ($\Delta \phi_{\text{L}} = 1/\sqrt{N_{\text{OPA}} (N_{\text{OPA}} + 2)}$) with $N_{\text{OPA}} = 2 \sinh^2 g$ being the total mean photon number inside the interferometer. However, $N_{\text{OPA}}$ is small because in this scheme the photon number is only dependent on the OPA strength $g$ which is of order of $3$ available currently \cite{agafonov2010,hudelist2014quantum}. Moreover, this sensitivity has been also discussed in Refs. \cite{anisimov2010,su11}. Although it beats the scaling of $1/N_{\mathrm{Tot}}$ with vacuum state input, one can find that the phase sensitivity approaches but does not reach below the ultimate quantum limit, the QCRB, as depicted in Fig. \ref{phi}(a). It is worthy noting that it does not surpass the corresponding HL for the optimal sensitivity of phase estimation under the conditions of these two inputs of one-coherent state and coherent $\otimes$ squeezed-vacuum state as depicted in Figs. \ref{phi}(b) and \ref{phi}(c), respectively.

\begin{figure}[tb!]
\includegraphics[scale=0.58]{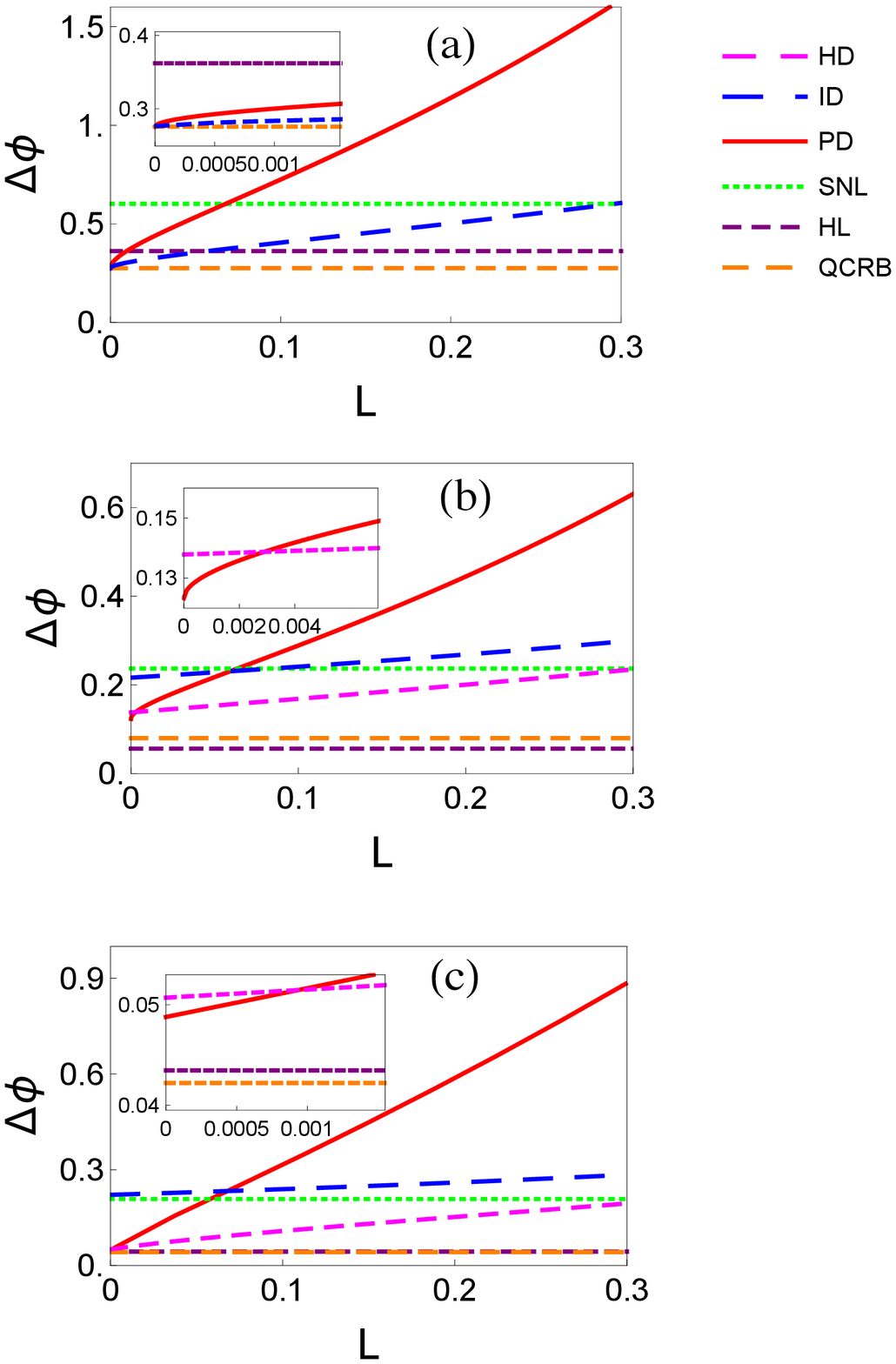}
\caption{The optimal phase sensitivity with the PD
versus the loss $L$ with various inputs of (a) vacuum state ($|
\alpha_0| = 0$ and $r = 0$); (b) only one coherent state ($|\alpha%
_0| = 2$ and $r = 0$); (c) coherent and squeezed vacuum state ($|
\alpha_0| = 2$ and $r = 1$). The dashed-magenta, dashed-blue, solid-red, dot-dashed-blue, dashed-purple, and dashed-orange lines correspond to the HD, the ID, the PD, the SNL, the HL, and the QCRB, respectively. It is worthy pointing out that the HD is missing in Fig. \ref{L}(a) due to its performance being always bad \cite{fig004}. Parameter used is: $g = 1$.}
\label{L}
\end{figure}

Corresponding to Fig.~\ref{phi}(a) [\ref{phi}(b) and \ref{phi}(c)], the optimal
phase sensitivity with the PD as a function of loss is shown in Fig.~\ref{L}(a) [\ref{L}(b) and \ref{L}(c)]. In Fig. \ref{L}(a), one can see that although $\Delta \phi _{\mathrm{L}}$ becomes worse as the loss increases, the optimal sensitivity of phase estimation could still beat the SNL when $L<0.07 = L_{\text{cri}}$ where we write this critical loss as $L_{\text{cri}}$ (the subscript cri stands for critical). It is worthy pointing out that $L<0.07$ indicates the lose being less than $7\%$ of photon number on each mode $a_1$ and $b_1$ between the two OPAs. In the situation of vacuum input considered here, it is $N_{\mathrm{Tot}} \equiv \langle \hat{a}_1^{\dagger} \hat{a}_1 + \hat{b}_1^{\dagger} \hat{b}_1\rangle = 2 \sinh^2 g \approx 2.8$ for the mean total photon number inside the SU(1,1) interferometer. Therefore, it is easily found that $L<0.07$ corresponds the photon number of loss being $0.07 N_{\text{Tot}} \approx 0.2$. 

Moreover, it is worthy pointing out that in Figs.~\ref{L}%
(b) and~\ref{L}(c), the critical losses are $L_{\text{cri}} \approx 0.06 $ (for one-coherent-state input with $|\alpha_0| = 2$ and $g=1$) and $L_{\text{cri}} \approx 0.05 $ (for coherent and squeezed-state input with $|\alpha_0| = 2$, $r=1$, and $g=1$), respectively. One can easily find that they are different for $L_{\text{cri}}$ in Figs.~\ref{L}%
(a), \ref{L}(b), and~\ref{L}(c). It is of significance to understand the behaviors of $L_{\text{cri}}$ since $L_{\text{cri}}$ determines whether it achieves a phase sensitivity below the SNL which can not be attainable by classical schemes. Naturally, a new question arises from that how $L_{\text{cri}}$ behaves with different experimental configurations. 

\begin{figure}[tb!]
\includegraphics[scale=0.645]{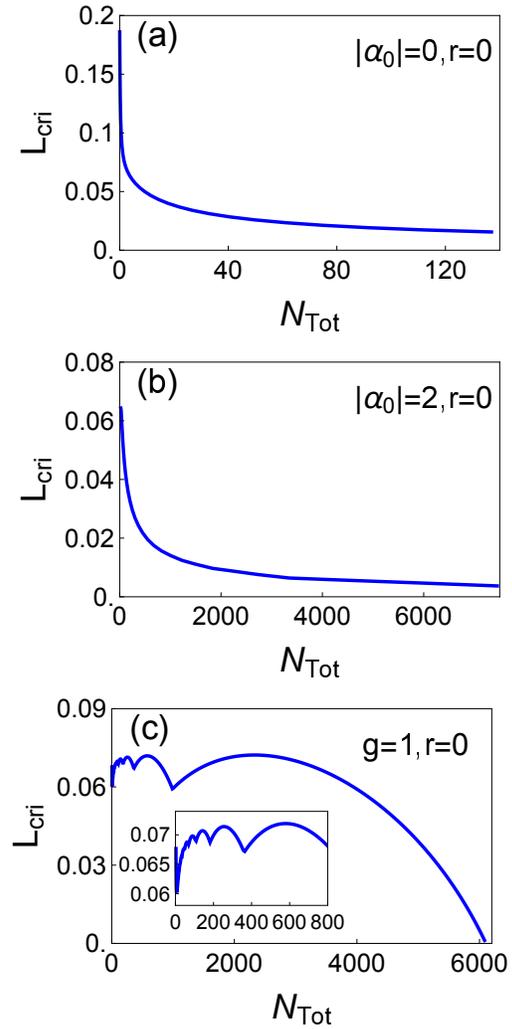}
\caption{The loss $L_{\text{cri}}$ as a function of $N_{\text{Tot}}$ where only when $L< L_{\text{cri}}$ the $\Delta \phi_{\text{L}}$ reaches below the SNL, with the PD for various cases: (a) vacuum input (varying $g$), (b) one-coherent-state input with a fixed amplitude of $|\alpha_0| = 2$ (varying $g \in [1,4]$), and (c) one-coherent-state input with a fixed value of $g = 1$ (varying $|\alpha_0| \in [0,100]$). Note that $N_{\text{Tot}} = (N_{\text{OPA}} + 1) N_{\text{in}} + N_{\text{OPA}}$.}
\label{Lsnl}
\end{figure}

To investigate it, we consider three different situations: (1) vacuum state input in Fig. \ref{Lsnl}(a), (2) one-coherent-state input with a fixed $|\alpha_0| = 2$ in Fig. \ref{Lsnl}(b), and (3) one-coherent-state input with a fixed value of $g = 1$ in Fig. \ref{Lsnl}(c)]. It is found that the critical loss $L_{\text{cri}}$ which degrades $\Delta \phi$ to approach the SNL is closely related to the detail configurations, such as, input states and OPA strengths. For instance, considering the vacuum input ($|\alpha| = 0$ and $r=0$), let the optimal phase sensitivity $\Delta \phi _{\text{L}}|_{\phi = \phi_{\text{opt}}} = \Delta \phi_{\text{SNL}}$, one can then obtain the relation between the loss $L_{\text{cri}}$ and $g$ which is rather complicated and is not reported here. To make it clear, we plot the $L_{\text{cri}}$ versus $N_{\text{Tot}}$ in Fig. \ref{Lsnl}(a). It shows that the $L_{\text{cri}}$ decreases with the increase of $g$ which indicates that with $N_{\text{Tot}}$ increasing, it becomes more easily to degrade to the SNL in the presence of loss. Moreover, in Fig. \ref{Lsnl}(b), the coherent state is fixed by $|\alpha_0| = 2$ while we adjust $N_{\text{Tot}}$ by varying $g$. It is easily found that $L_{\text{cri}}$ decreases with $N_{\text{Tot}}$ increasing. Furthermore, for comparison, in Fig. \ref{Lsnl}(c), $g$ is fixed while we vary $|\alpha_0|$ to set $N_{\text{Tot}}$. In such a condition, when $N_{\text{Tot}} > 4000$, $L_{\text{cri}}$ decreases gradually with the increase of $N_{\text{Tot}}$. However, $L_{\text{cri}}$ fluctuates around $0.06$ when $N_{\text{Tot}} < 4000$ which indicates that the phase sensitivity can beat the SNL with a certain loss rate even within a wide range of input intensity.

\section{Comparison among different detections in the presence of loss}

In a lossy traditional SU(2) interferometer, practical measurements have been compared including the HD, the ID and the PD \cite{gard2017,oh2017}. Naturally, it is also necessary to compare
the performances among the HD, the ID, and the PD in a lossy SU(1,1) interferometer since the SU(1,1) has been realized experimentally \cite{jing2011}. 

It is worthy noting that the traditional SU(2) interferometer measures the difference in the intensities of the two output modes generally whereas the ID measures the total number of photons at the output in an SU(1,1) interferometer. This is due to for an SU(1,1) interferometer, the intensity difference between the two output modes is a conserved quantity \cite{leonhardt1994} and provides no information about $\phi$ since photons are always created or eliminated in pairs by the parametric processes.

Similar to method in Ref. \cite{marino12}, one can obtain the sensitivity of phase variance with coherent and squeezed vacuum state input in an SU(1,1) interferometer with the ID which is in a rather complicated expression and is shown in Appendix \ref{sec:ID}. Meanwhile, in Ref.~\cite{li2014phase}, it
presents the sensitivity of phase estimation with a coherent and
squeezed vacuum input state in the presence of loss with the HD,
\begin{equation}
\Delta \phi _{\mathrm{L}}^{\mathrm{H}}=\frac{1}{e^{r}G_{\mathrm{OPA}}\sqrt{N_{\alpha }}}[ 1+\frac{L}{
(1-L)}\frac{e^{2r}N_{\mathrm{Tot}}}{N_{\alpha}+N_{\mathrm{s}}}] ^{1/2},
\end{equation}%
where the superscript H denotes the HD.

\begin{figure}[tb!]
\includegraphics[scale=.58]{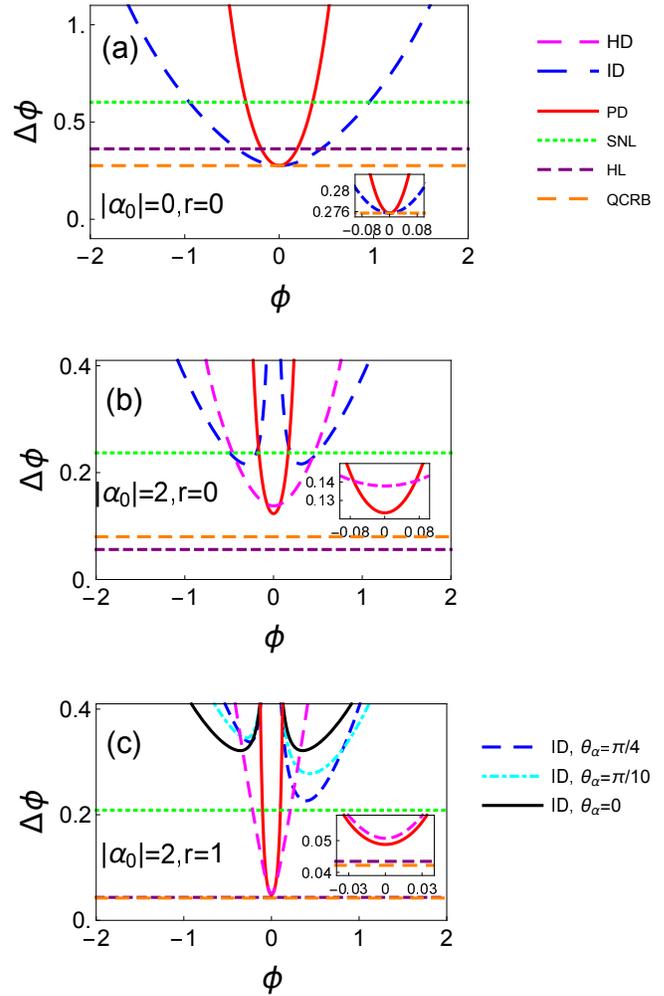}
\caption{The phase sensitivity with parity detection
versus $\phi$ for different detections with various inputs of (a) vacuum state ($|\protect%
\alpha_0| = 0$ and $r = 0$); (b) only one coherent state ($|\protect\alpha%
_0| = 2$ and $r = 0$); (c) coherent and squeezed vacuum state ($|\protect%
\alpha_0| = 2$ and $r = 1$). The dashed-magenta, dashed-blue, solid-red, dot-dashed-blue, dashed-purple, and dashed-orange lines correspond to the HD, the ID, the PD, the SNL, the HL, and the QCRB, respectively. Similar to the Fig. \ref{L}(a), the HD is also missing in Fig. \ref{parityhomo}(a) due to its performance being always bad \cite{fig004}. Moreover, it is worthy noting that in Fig. \ref{parityhomo}(c), it plots the phase sensitivities with the ID for different phases of coherent input state. This is owing to that the optimal phase of $\theta_\alpha$ is not quite clear with the ID which determines the optimal phase sensitivity to be achieved whereas it is optimal phase $\theta_\alpha = 0$ for both the HD and the PD. In Fig. \ref{parityhomo}(c), solid-black, dot-dashed-cyan, and dashed-blue lines denote the ID with $\theta_\alpha = 0, \pi/10,$ and $\pi/4$, respectively. Parameter used is: $g = 1$.}
\label{parityhomo}
\end{figure}

The comparison is performed among the HD, the ID, and the PD for the optimal phase sensitivity as a function of loss $L$ $ (L_1=L_2=L)$ as shown in Fig.~\ref{L}: (a) vacuum input, (b) one-coherent-state input, and (c) coherent and squeezed-state input, where the dashed-magenta, dashed-blue, solid-red, dotted-green, purple-dashed, and dashed-orange lines stand for the HD, the ID, the PD, the SNL, and the HL, respectively. In Fig.~\ref{L}(a), we see that with vacuum state input ($|\alpha_0|=0$) the ID has a better performance than the PD in the
presence of loss. With small loss $L < 0.07$ the PD can still surpass the
SNL while the ID can beat the SNL even with large loss $L < 0.3$. Although
the PD has a phase sensitivity as same as the ID in the ideal case as depicted in the zoom figure in Fig.~\ref{L}(a), the ID is more optimal than
the PD in practical experiments in the presence of loss. Moreover, it is worthy pointing out that under the condition of vacuum state input, the HD has a phase sensitivity which is always bad \cite{fig004}. Therefore, the HD is missing in Fig.~\ref{L}(a).

In Fig.~\ref{L}(b), it plots the optimal phase sensitivities for various detections with one-coherent-state input. It is easily found that although without loss the PD has the best phase sensitivity as shown in the corresponding inset figure, the HD becomes the optimal detection scheme with the loss increasing. Moreover, even with large loss $L<0.3$ the HD can still beat the SNL which is a robust scheme to the loss. Furthermore, it is worthy noting that with one-coherent-state input the ID has a optimal phase sensitivity which is around the SNL and is worst than the ones of the HD and ID. To explain this, we plot the phase sensitivities of the HD, the ID, and the PD versus $\phi$ in the absence of loss as shown in Fig.~\ref{parityhomo}(b). One can easily find that the ID has a phase sensitivity which diverges at $\phi=0$ whereas $\phi=0$ is the optimal phase point for both the HD and the PD. Therefore, the ID has a worse optimal phase sensitivity which is not suitable for the case of one-coherent-state input in a nonlinear interferometer. However, the ID has a phase sensitivity which degrades a little with the increase of loss as depicted in Fig. \ref{L}(b). Therefore, even with a worse optimal phase sensitivity, the ID is robust to the loss.

Fig. \ref{L}(c) presents the optimal phase sensitivities of the HD, the ID, and the PD versus loss $L$ with the input of coherent and squeezed state. In the absence of loss, the optimal phase sensitivities of HD and PD are close to the HL whereas it is around the SNL for the optimal phase sensitivity with the ID. Similar to the case of one-coherent-state input, the PD has the best optimal phase sensitivity without loss with coherent and squeezed-state input as shown in the zoom figure in Fig. \ref{L}(c). However, it degrades sharply for the optimal phase sensitivity of the PD with the loss increasing while the ones of the ID and HD degrade a little.

Corresponding to Fig. \ref{L}(c), it plots the phase sensitivities in the absence of loss as a function of $\phi$ with coherent and squeezed-state input for various detection schemes in Fig. \ref{parityhomo}(c). It is worthy pointing out that it diverges at $\phi = 0$ for the phase sensitivity of the ID which is similar to the case of one-coherent-state input. Moreover, it is found that the phase sensitivity is related the phase of coherent input state $\theta_\alpha$ where $\theta_s$ has been set to zero for the phase of the squeezed input state. Therefore, it also plots the phase sensitivities for various $\theta_\alpha = 0, \pi/10,$ and $\pi/4$ in Fig. \ref{parityhomo}(c). It is worthy noting that when $\theta_\alpha = 0$, it is symmetrical for the phase sensitivity with the ID along with the line of $\phi=0$ whereas it is not symmetrical any more when $\theta_\alpha = \pi/10$ and $\pi/4$. However, if $\theta_\alpha = \pi/4$, it has a better optimal phase sensitivity than the cases of $\theta_\alpha = 0$ and $ \pi/10$.

Although the ID achieves its optimal phase sensitivity as well as the scaling of phase sensitivity as shown in Fig. \ref{L}(c), it is close to the SNL which looks like not consistent with the conclusion of beating the SNL as reported in previous works \cite{plick2010,marino12}. In fact, the results do not contradict each other. In Refs. \cite{plick2010,marino12}, they presented that the ID is proven to be an efficient detection scheme in which the input is in a two-coherent state whereas the input is in a coherent and squeezed state in our scheme. Therefore, the input considered here is different from the previous works which leads to a medium performance of phase sensitivity with the ID.  

To verify this, it is investigated for the case of two-equal-coherent state as input. Similarly, the comparison is performed among these three detections in an SU(1,1) interferometer in the following content. According to Ref. \cite{marino12}, it can be found that the optimal sensitivity of phase estimation with the ID with considering loss with two-equal-coherent-state input $|i \alpha_0/\sqrt{2}\rangle \otimes |\alpha_0 / \sqrt{2}\rangle$ is given by 
\begin{align}
\Delta \phi_{\text{L,twocoh}}^{\text{I}} =& \{ \frac{1}{4 |\alpha_0|^2 \sinh^2 g \cosh^2 g} [1 + \frac{L}{1- L} (1 + 2 \sinh^2 g)]  \nonumber \\
&+ \frac{L}{(1-L)^2} \frac{1+ L (\sinh^2 g + \cosh^2 g)}{ 4 |\alpha_0|^2 \cosh^2 g}\}^{1/2},
\end{align}
where for convenience, we have used $g$, $1-L$, $i\alpha_0/\sqrt{2}$ and $\alpha_0/\sqrt{2}$ to replace the $s$, $\eta_1$, $\alpha$ and $\beta$ in the expression of Eq. (23) in Ref. \cite{marino12}, respectively. Similar to the method as discussed in Ref. \cite{li2014phase}, one can obtain the optimal phase sensitivity with the HD with two-equal-coherent-state input as
\begin{align}
\Delta \phi_{\text{L,twocoh}}^{\text{H}} =& \sqrt{\frac{ L \cosh (2 g)+1-L}{ 1-L }}\frac{1}{ \sqrt{2}|\alpha_0|  \cosh^2 g (\tanh
   g+1)} \nonumber \\
=& \sqrt{\frac{ L \cosh (2 g)+1-L}{ 1-L }}\times \nonumber \\
&\frac{\sqrt{2}}{ |\alpha_0| (\sqrt{N_{\text{OPA}} (N_{\text{OPA}} + 2)} + N_{\text{OPA}} + 2) },
\end{align}
where we have used $N_{\text{OPA}} \equiv 2 \sinh^2 g$ which means the spontaneous photon number generated by the fisrt OPA when the input state is in a vacuum state.

\begin{figure}[tb!]
\includegraphics[scale=.6]{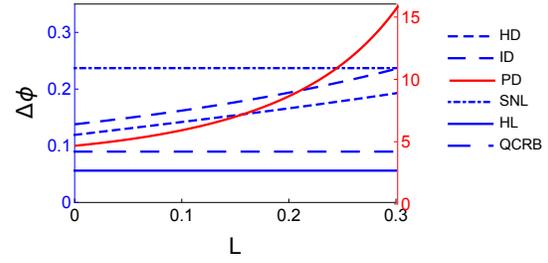}
\caption{The optimal phase sensitivity $\Delta\phi$ of the
SU(1,1) interferometer as a function of loss $L$ for different detection schemes (HD, ID, and PD) with two-equal-coherent-state input $|i \alpha_0 /\sqrt{2}\rangle \otimes |\alpha_0/\sqrt{2}\rangle$. The short-dashed-blue, long-dashed-blue, solid-red, dot-dashed-blue, and solid-blue lines correspond to the HD, the ID, the PD, the SNL, and the HL, respectively. It is worthy pointing out that the four blue lines are related to the blue scaling on the left side of frame while the red line corresponds to the red scaling on the right side one. Parameters used are: $|\alpha_0| = 2$ and $g
= 1$.}
\label{fig007}
\end{figure}

In Fig. \ref{fig007}, we plot the optimal sensitivities of phase estimation as a function of loss with two-equal-coherent-state input $|i \alpha_0/\sqrt{2}\rangle \otimes |\alpha_0 / \sqrt{2}\rangle$ for various detection schemes. The short-dashed-blue, long-dashed-blue, solid-red, dot-dashed-blue, solid-blue, and dashed-orange lines correspond to the HD, the ID, the PD, the SNL, the HL, and the QCRB, respectively. It is worthy pointing out that the five blue lines are related to the blue scaling on the left side of frame while the red line corresponds to the red scaling on the right side one. One can easily find that the HD has the best optimal sensitivity of phase estimation among those three detection schemes while the PD has a phase sensitivity which is worst under the condition of two-equal-coherent-state input. Moreover, it is worthy noting that although the ID has a medium performance, it still surpasses the SNL even with loss of $L = 30\%$ which is different from the cases of one-coherent-state and coherent $\otimes$ squeezed-state inputs. Furthermore, it is easily to find that both the HD and the ID are robust to the loss where the optimal phase sensitivities degrade a little with the increase of the loss as depicted in Fig. \ref{fig007}.

\begin{figure}[tb!]
\includegraphics[scale=.72]{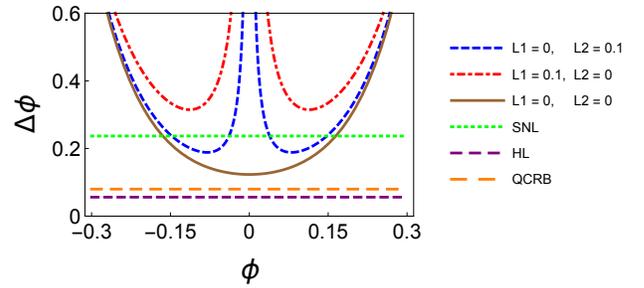}
\caption{The phase sensitivity $\Delta\protect\phi$ of the
SU(1,1) interferometer as a function of $\protect\phi$ with the loss on the
phase sensing arm (dot-dashed-red line) and the free arm (dashed-blue line),
respectively. The solid-brown line presents the ideal case. The dashed-green, dashed-purple, and dashed-orange lines denote the SNL, the HL, and the QCRB, respectively. Parameters used are: $|\protect\alpha_0| = 2$, $r = 0$, and $g
= 1$.}
\label{figcohl1l2}
\end{figure}

It has been studied for the case of the same value of loss ($L_1 = L_2$) on the two arms between the two OPAs in the previous section. Here we will discuss the phase
sensitivity in the presence of different values of loss, $L_1 \neq L_2$,
where the result is shown in Appendix~\ref{sec:cohl1l2}. In Fig.~\ref{figcohl1l2},
we plot the sensitivities as a function of $\phi$ in the presence of loss
only on the phase sensing arm (dot-dashed-red line: $L_1 = 0.1, L_2 = 0$)
or the free arm (dashed-blue line: $L_1 = 0, L_2 = 0.1$), respectively. It
is shown that the sensitivity reduction by the loss on the phase sensing arm
is larger. This is due to the fact that the loss on the phase sensing arm degrades the
phase information directly which has a significant impact on the sensitivity
of phase estimation. Therefore, we should pay much more attention on the
loss on the arm experiencing phase shift in experiments.

\section{Conclusion}
In conclusion we study the effects of loss on the phase sensitivity in an
SU(1,1) interferometer with the PD. The effects of loss have a significant role on the sensitivity of phase estimation with the PD while the PD still serves as an optimal strategy when considering the phase fluctuations~\cite{Bardhan:2013}. This is due to photon loss distorting the parity of photon number for the PD to estimate the phase variance. In the presence of small loss, the PD still surpasses the SNL. Moreover, we also compare the performance among the HD, the ID, and the PD in a lossy nonlinear interferometer with various state inputs. It is found that when the input is in a vacuum state, the ID is the optimal detection in the presence of loss while the HD is the optimal measurement when the input is in a one-coherent state, coherent $\otimes$ squeezed state, or two-equal-coherent state in the presence of loss.

\section{Acknowledge}
Dong Li is supported by the Science Challenge Project under Grant No. TZ2018003-3. Chun-Hua Yuan is supported by the National Natural Science Foundation of China under Grant No. 11474095 and the Fundamental Research Funds for the Central Universities. Yao Yao is supported by the National Natural Science Foundation of China under Grant No. 11605166. Weiping Zhang acknowledges support from National Key Research and Development
Program of China under Grant No. 2016YFA0302001,
the National Natural Science Foundation of China
(Grants No. 11654005, No. 11234003, and No.
11129402), and the Science and Technology Commission
of Shanghai Municipality (Grant No. 16DZ2260200).

\begin{thebibliography}{99}
\bibitem{su11} The MZI with two-mode squeezed-vacuum state input has a similar phase sensitivity with the SU(1,1) interferometer with vacuum state input.
\bibitem{fig004}
The definition of observable, quadrature, we used here is
$\hat{X}_{b2} \equiv (\hat{b}_2^{\dagger} + \hat{b}_2)/\sqrt{2}$. For vacuum inputs, the signal of homodyne
detection is always a constant $\langle \hat{X}_{b2} \rangle = 0$, whereas the variance of signal of homodyne detection is given by $\langle \Delta \hat{X}_{b2} \rangle \equiv ( \langle \hat{X}_{b2}^2 \rangle - \langle \hat{X}_{b2}\rangle^2)^{1/2} = \sqrt{2 \cosh ^2 g \sinh^2 g (1-\cos \phi) + 1/2} $, where $g$ is the OPA strength and $\phi$ is the phase shift. According to the error propagation formula $\Delta \phi_{\text{H}} = \langle \Delta \hat{X}_{b2} \rangle /|\partial \langle \hat{X}_{b2} \rangle/ \partial \phi|$, one finds that the phase sensitivity is always bad and not available in such a case.
\end{thebibliography}

\onecolumn
\section{Appendix}

\subsection{Model}
\label{model}
We review the model as discussed in Ref.~\cite{li2016phase}. The schematic of an SU(1,1) interferometer is shown in Fig.~\ref{fig1}
where the OPAs take the place of the 50-50 beam splitters in a traditional
MZI. Here a coherent light mixed with a squeezed vacuum state is considered
as the input. The annihilation (creation) operators of the two modes are
denoted by $\hat{a}$ $(\hat{a}^{\dag })$ and $\hat{b}$ $(\hat{b}^{\dag })$.
The propagation of the beams through the SU(1,1) interferometer is described
as follow: after the first OPA, one output is retained as a reference, while
the other one experiences a phase shift; after the beams recombine in the
second OPA with the reference field, the output lights are dependent on the
phase difference $\phi$ between the two modes.

For simplicity, we consider the phase space to describe the propagation. The
Wigner function of the input state, a product state $|\alpha
_{0}\rangle \otimes |0,\xi =re^{i\theta _{s}}\rangle $, with coherent light
amplitude $\alpha _{0}=|\alpha _{0}|e^{i\theta _{\alpha }}$, is written as%
\begin{equation}
W_{\mathrm{in}}(\alpha _{i},\alpha _{0};\beta _{i},r)=W_{|\alpha _{0}\rangle
}(\alpha _{i},\alpha _{0})W_{|0,\xi \rangle }(\beta _{i},r),
\end{equation}%
where the Wigner functions of coherent and squeezed vacuum states are given
by \cite{walls2007quantum}%
\begin{align}
W_{|\alpha _{0}\rangle }(\alpha _{i},\alpha _{0}) &=\frac{2}{\pi }%
e^{-2|\alpha _{i}-\alpha _{0}|^{2}}, \\
W_{|0,\xi \rangle }(\beta _{i},r) &=\frac{2}{\pi }e^{-2|\beta
_{i}|^{2}\cosh 2r+(\beta _{i}^{2}+\beta _{i}^{\ast 2})\sinh 2r},
\end{align}%
respectively, with $\theta _{s}=0$ by appropriately fixing the irrelevant
absolute phase $\theta _{\alpha }$, and $\beta _{i}^{\ast }$ is the
conjugate of $\beta _{i}$.

After propagation through the SU(1,1) interferometer the output Wigner
function is found to be%
\begin{equation}
W_{\mathrm{out}}(\alpha _{f},\beta _{f})=W_{\mathrm{in}}[\alpha _{i}(\alpha
_{f},\beta _{f}),\beta _{i}(\alpha _{f},\beta _{f})],
\end{equation}%
where $\alpha _{i},\beta _{i},\alpha _{f}$ and $\beta _{f}$ are the complex
amplitudes of the beams in the mode $\hat{a}_{0},\hat{b}_{0},\hat{a}_{2}$,
and $\hat{b}_{2}$, respectively. Generally, propagations through the first
OPA, phase shift, and second OPA are given by%
\begin{align}
T_{\mathrm{OPA1}}=\left(
\begin{array}{cc}
u_{1} & v_{1} \\
v_{1}^{\ast } & u_{1}%
\end{array}%
\right), \quad \quad
 T_{\phi }=\left(
\begin{array}{cc}
e^{i\phi } & 0 \\
0 & 1%
\end{array}%
\right), \quad \quad
 T_{\mathrm{OPA2}} =\left(
\begin{array}{cc}
u_{2} & v_{2} \\
v_{2}^{\ast } & u_{2}%
\end{array}%
\right),
\label{u1}
\end{align}%
with $u_{k}=\cosh g_{k}, v_{k}=e^{i\theta _{k}}\sinh g_{k}$, and $v_{k}^{\ast}$ being the conjugate of $v_{k}$ $ (k = 1,2)$, where $\theta _{1}$ $(\theta _{2})$ and $g_{1}$ $(g_{2})$ are the phase shift and parametric strength in the OPA
process 1 (2), respectively, see, for example Ref.~\cite{xu2014optical}.
Therefore, the nonlinear interferometer is described by $T=T_{%
\mathrm{OPA2}}T_{\phi }T_{\mathrm{OPA1}}$. Thus, the relation
between variables is given by%
\begin{equation}
\left(
\begin{array}{c}
\alpha _{i} \\
\beta _{i}^{\ast }%
\end{array}%
\right) =T^{-1}\left(
\begin{array}{c}
\alpha _{f} \\
\beta _{f}^{\ast }%
\end{array}%
\right) .
\end{equation}%
More specifically, we assume that the first OPA and the second one have a $%
\pi $ phase difference (particularly $\theta _{1}=0$ and $\theta _{2}=\pi $)
and same parametric strength ($g_{1}=g_{2}=g$). In such a case, the second
OPA will undo what the first one does (namely $\hat{a}_{2}=\hat{a}_{0}$ and $%
\hat{b}_{2}=\hat{b}_{0}$) if phase shift $\phi =0$, which we call the
balanced situation.

In the case of the balanced situation, the relation between the input and
output variables is found to be

\[
T^{-1}=\left(
\begin{array}{cc}
G & R \\
-R & H%
\end{array}%
\right) ,
\]%
where $G=A-iB\cosh (2g),H=A+iB\cosh (2g)$ and $R=-iB\sinh (2g)$ with $A=\cos
(\phi /2)e^{-i\phi /2}$ and $B=\sin (\phi /2)e^{-i\phi /2}$. The
output Wigner function of the SU(1,1) interferometer is then obtained,%
\begin{align}
W_{\mathrm{out}}(\alpha _{f},\beta _{f}) =&\frac{4}{\pi ^{2}}e^{-2|G\alpha
_{f}+R\beta _{f}^{\ast }-\alpha _{0}|^{2}-2|-R\alpha _{f}+H\beta _{f}^{\ast
}|^{2}\cosh (2r)} e^{2\mathrm{Re}[(-R\alpha
_{f}+H\beta _{f}^{\ast })^{2}]\sinh (2r)}.  \label{wigner}
\end{align}

Alternatively, according
to Ref.~\cite{Wigner77}, the parity signal is given by%
\begin{equation}
\langle \hat{\Pi}_{b}\rangle =\frac{\pi }{2}\int W_{\mathrm{out}}(\alpha
_{f},0)d^{2}\alpha _{f},  \label{signal}
\end{equation}%

\subsection{Parity detection signal in ideal case}
\label{sec:signal} According to Eqs.~(\ref{wigner}) and~(\ref{signal}),
the detection signal $\langle\hat{\Pi}_{b}\rangle$ is worked out as%
\begin{equation}
\langle\hat{\Pi}_{b}\rangle=\frac{1}{\sqrt{x_{1}}}e^{-x_{2}/x_{3}},
\end{equation}
where
\begin{align}
x_{1} = &e^{-2r} (e^{2r}+1)^{2} [8 \sinh^{4}(2g) (\cos(2\phi)-\cos \phi) +
4\cosh (4g)   + 3 \cosh(8g) -7 ] + 64,  \nonumber \\
x_{2}=&4|\alpha|^{2}\sinh^{2}(2g)\{8\cosh (4g)\cos(2\theta)\sin^{4}( \phi/2) -8\cosh(2g)\sin(2\theta )\sin\phi(\cos\phi-1)  \nonumber \\
&+8e^{4r}[\cos\theta\sin\phi-2\cosh(2g)\sin\theta\sin ^{2}( \phi/2) ]^{2}+32e^{2r}\sinh^{2}(2g)\sin^{4}( \phi/2)  \nonumber \\
&+8\cosh(4g)\sin^{4}( \phi/2) -8\cos^{2}\theta \cos\phi +[3\cos(2\theta)-1]\cos(2\phi)+\cos(2\theta)+5\},  \nonumber \\
x_{3} = &(e^{2r}+1)^{2}[8 \cosh(8g)\sin^{4}(\phi/2) + 8\cosh(4g)\sin^{2} \phi + 4 \cos \phi + 3 \cos (2 \phi) - 7] + 64 e^{2r}.
\end{align}
Assuming that $\phi=0,$ the signal is then simplified to
\begin{align}
\langle\hat{\Pi}_{b}\rangle|_{\phi=0}=1.
\end{align}
This is due to the second OPA would undo what the first one did if $\phi=0$%
. Meanwhile the output field is the same as the input. Thus the output in
mode $b$ is the one-mode squeezed vacuum. The measurement signal is one with
the one-mode squeezed vacuum input causing only even number distribution in
the Fock basis with $|0,\xi=re^{i\phi_{s}}\rangle=\sqrt{1/\cosh r}%
\sum_{n=0}^{\infty }(\sqrt{(2n)!}/n)(1/2)^{n}[\exp(i\phi_{s})\tanh
r]^{n}|2n\rangle$~\cite{barnett2002methods}.

\subsection{Heisenberg limit and shot-noise limit}
\label{sec:hl}
Here, we focus on the HL in an SU(1,1) interferometer. According to Ref.~\cite{marino12}, the HL corresponds to the total number of photons $N_{\mathrm{Tot}} (\equiv \langle \hat{a}_1^{\dagger} \hat{a}_1 + \hat{b}_1^{\dagger} \hat{b}_1\rangle)$ inside the SU(1,1) interferometer not the input photon number as the traditional MZI. This is due to amplification of the phase sensing photon number by the first OPA. Then the corresponding HL is given by 
\begin{equation}
\Delta \phi _{\mathrm{HL}}=\frac{1}{N_{\mathrm{Tot}}},  \label{HL}
\end{equation}%
where the subscript HL represents Heisenberg limit. According to Ref.~\cite{li2014phase} the
total inside photon number is given by%
\begin{equation}
N_{\mathrm{Tot}}=(N_{\mathrm{OPA}}+1)N_{\mathrm{in}}+N_{\mathrm{OPA}},
\label{HL1}
\end{equation}%
where $N_{\mathrm{in}}=N_{\alpha }+N_{s}$. The first term of equation (\ref%
{HL1}) on the right-hand side, $(N_{\mathrm{OPA}}+1)N_{\mathrm{in}}$,
results from the amplification process of the input photon number and the
second one is related to the spontaneous process. Thus the total inside
photon number $N_{\mathrm{Tot}}$ corresponds to not only the OPA strength
but also the input photon number. Then the corresponding SNL is written as%
\begin{equation}
\Delta \phi _{\mathrm{SNL}}=\frac{1}{\sqrt{N_{\mathrm{Tot}}}},  \label{SNL}
\end{equation}%
where the subscript SNL denotes the shot-noise limit.

\subsection{Quantum Cram\'er-Rao bound}
\label{sec:qcrb}
According to Ref.~\cite{li2016phase}, one can easily find the QCRB of an SU(1,1) interferometer with different input states which are shown in the table as follows.

\begin{table*}[!h]
\tabcolsep 3mm
\doublerulesep 8mm
\caption{The quantum Cram\'er-Rao bound (QCRB) of an SU(1,1)
interferometer with different input states.}
\begin{center}
\renewcommand\arraystretch{2}
\begin{tabular}{p{0.2\textwidth}p{0.551\textwidth}}
\hline
Input states & QCRB\\\hline
$|0\rangle\otimes|0\rangle$  &  $1/\mathcal{K}^{1/2}$ \\
$|\alpha_{0}\rangle\otimes|0\rangle$  & $1/[\mathcal{K}(2 N_{\alpha} +1)+2 N_{\alpha}(N_{\text{OPA}} + 2)]^{1/2}$\\
$|\frac{i\alpha_{0}}{\sqrt{2}}\rangle \otimes|\frac{\alpha_{0}}{\sqrt{2}}\rangle$  & $1/\{2N_{\alpha }[(N_{\text{OPA}}+1)\sqrt{\mathcal{K}}+\mathcal{K}+1]+\mathcal{K}\}^{1/2}$\\
$|\alpha_{0}\rangle\otimes|0,\xi\rangle$  &  $1/[2N_{\alpha }(N_{\text{OPA}}+2)+N_{\text{OPA}}^{2}\sinh ^{2}(2r)/2+\mathcal{K}(2 N_{\alpha } \cosh r e^{r}+\cosh ^{2}r)]^{1/2}$ \\\hline
\end{tabular}\\
{where $\mathcal{K}=N_{\text{OPA}}(N_{\text{OPA}}+2)$. Row 1: vacuum input state;
Row 2: one-coherent input state; Row 3: two-coherent input state; Row 4: coherent mixed with squeezed-vacuum input state.\label{Tab001}}
\end{center}

\end{table*}

\subsection{Parity detection signal in the presence of loss}
\label{sec:signalloss} According to Eqs.~(\ref{outwloss}),~(\ref%
{inputoutloss}), and~(\ref{wloss}), the parity signal in the presence of
loss is given by%
\begin{align}
\langle \hat{\Pi}_b^{\mathrm{loss}} \rangle = \frac{8 }{\sqrt{y_1}}e^{-\frac{%
y_2}{y_3}},
\end{align}
where
\begin{align}
y_1 = &e^{-2 r} (-4 \cosh (4 g) (-2 (5 L^2-2 L+1) e^{2 r}+(L-1)^2 (e^{2
r}+1)^2 \cos (2 \phi )  +(L-1)^2 (-e^{4 r})-(L-1)^2)+16 (L-1)^2 e^{2 r} \cosh (6 g) \cos
\phi  \nonumber \\
&+8 (L-1)^2 e^{4 r} \cosh (6 g) \cos \phi-8 (L-1)^2 e^{2 r} \cosh
(8 g) \cos \phi  -4 (L-1)^2 e^{4 r} \cosh (8 g) \cos \phi+2 (L-1)^2 e^{2 r} \cosh
(8 g) \cos (2 \phi )  \nonumber \\
&+(L-1)^2 e^{4 r} \cosh (8 g) \cos (2 \phi )+16 (L-1) e^{2 r}
\cosh (6 g) \cos \phi  +8 (L-1) e^{4 r} \cosh (6 g) \cos \phi+8 (1-L) L \cosh (2 g)
((e^{2 r}+1)^2 \cos \phi  \nonumber \\
&-2 e^{2r}+7 e^{4 r}+7)-16 (L-1)^2 e^{2 r} \cosh (6 g)-8 (L-1)^2
e^{4 r} \cosh (6 g)  +6 (L-1)^2 e^{2 r} \cosh (8 g)+3 (L-1)^2 e^{4 r} \cosh (8 g)  \nonumber \\
&-16
(L-1) e^{2 r} \cosh (6 g)-8 (L-1) e^{4 r} \cosh (6 g)+8 (L-1)^2 \cosh (6 g) \cos \phi-4
(L-1)^2 \cosh (8 g) \cos \phi  +(L-1)^2 \cosh (8 g)\cos (2 \phi )\nonumber \\
&+8 (L-1) \cosh (6 g) \cos
\phi-8 (L-1)^2 \cosh (6 g)  +3 (L-1)^2 \cosh (8 g)-8 (L-1) \cosh (6 g)+8 (L-1)^2 e^{2 r} \cos
\phi  +4 (L-1)^2 e^{4 r} \cos \phi\nonumber \\
&+6 (L-1)^2 e^{2 r} \cos (2 \phi )
+3 (L-1)^2 e^{4 r} \cos (2 \phi )+82 (L-1)^2 e^{2 r}-7 (L-1)^2
e^{4 r}  +64 (L-1) e^{2r}+4 (L-1)^2 \cos \phi+3 (L-1)^2 \cos (2 \phi )
\nonumber \\
&-7 (L-1)^2+32 e^{2 r}),  \nonumber \\
y_2 = &16 |\alpha_0|^2 (1-L) \sinh ^2(2 g) \sin ^2(\frac{\phi }{2}) (2 (L-1)
(e^{2 r} (\cosh (4 g)-1)   \times(\cos \phi-1)+(\cosh (4 g)+1) (\cos \phi-1)  \nonumber \\
&-2 e^{4 r} (\cos \phi+1))+8 L e^{2 r} \cosh (2 g)),  \nonumber \\
y_3 = &-8 (L-1)^2 (e^{2 r}+1)^2 \sinh ^2(4 g) \cos \phi+2 e^{2 r} (8 (L-2) L
\sinh ^4(2 g) \cos (2 \phi )  +4 (L (5 L-2)+1) \cosh (4 g)-8 (L-1) L \cosh (6 g)  \nonumber \\
&+3 (L-1)^2 \cosh (8 g)+8 \sinh ^4(2 g) \cos (2 \phi )+L (41
L-50)+25)  +(L-1) e^{4 r} ((L-1) (8 \sinh ^4(2 g) \cos (2 \phi )+3 \cosh (8
g)-7)  \nonumber \\
&+4 (L-1) \cosh (4 g)-8 L \cosh (6 g))+8 (L-1) L \cosh (2 g)
 \times (4 (e^{2 r}+1)^2 \sinh ^2(2 g) \cos \phi+2 e^{2 r}-7 e^{4
r}-7)  \nonumber \\
&+(L-1) ((L-1) (8 \sinh ^4(2 g) \cos (2 \phi )+3 \cosh (8 g)-7)
+4 (L-1) \cosh (4 g)-8 L \cosh (6 g)).
\end{align}

\subsection{Phase sensitivity with the PD with one-coherent-state input with the same loss on the two arms}
\label{sec:cohl} If one-coherent state is injected, according to Eqs.~(%
\ref{wloss}) and~(\ref{sensiloss}), the phase sensitivity in the presence of
loss is worked out as%
\begin{eqnarray}
\Delta \phi_{\mathrm{L}} = \sqrt{\frac{z_1}{z_2}},
\end{eqnarray}
where
\begin{align}
z_1 = &1-\frac{4 \exp (\frac{8 |\alpha_0|^2 (L-1) \sinh ^2(2 g) \sin^2(\frac{%
\phi }{2})}{2 L \cosh (2 g)-(L-1) (-2 \sinh ^2(2g) \cos \phi+\cosh (4 g)+1)})%
}{((L-1) (-2 \sinh^2(2 g) \cos \phi+\cosh (4 g)+1)-2 L \cosh (2 g))^2}, \nonumber\\
z_2 = &256 \frac{k_1}{k_2}e^{k_3},
\end{align}
with
\begin{align}
k_1 = &(1-L)^2 \sinh ^4(2 g) \sin ^2\phi (-4 (|\alpha_0|^2+1) L \cosh (2 g)
+2 (1-L) \cosh (4 g) \cos \phi-(2-2 L) \cosh (4 g)  \nonumber \\
&-4 |\alpha _0|^2 (1-L)-2 (1-L) \cos \phi-3 (1-L)-L+1)^2, \nonumber\\
k_2 = &(4 (1-L) \sinh ^2(2 g) \cos (\phi )-4 L \cosh (2 g) -(2-2 L) \cosh (4 g)-3 (1-L)-L+1)^6, \nonumber\\
k_3 = &[4 |\alpha_0|^2 \sinh ^2(2 g) (-2 (1-L) \cos \phi -2 L+2)]/[4 (1-L)
 \sinh ^2(2 g) \cos \phi -4 L \cosh (2 g)-(2-2 L) \cosh (4
g)  \nonumber \\
&-3 (1-L)-L+1].
\end{align}

\subsection{Phase sensitivity via the ID in the presence of loss}
\label{sec:ID} If one-coherent state is injected with the loss of $L_1 = L_2 = L$, the phase sensitivity with the ID is found to be%
\begin{align}
\Delta \phi_{\text{L}}^{\text{I}} = &\{\text{csch}^4(2 g) [\csc
   ^2 (\frac{\phi}{2}) (2
   (|\alpha_0| ^2+1) L (1-L) \cosh
   (2 g)+2 |\alpha_0| ^2 (1-L)^2+L^2 \cosh (4
   g)-(2-L) L) \\\nonumber
   &+\sec
   ^2(\frac{\phi}{2}) ((2
   |\alpha_0| ^2+1) (1-L)^2 \cosh (8 g)+2
   (|\alpha_0| ^2+1) L (1-L) \cosh
   (6 g)+L^2 \cosh (4 g)-1)] \\\nonumber
   &-8
   (2 |\alpha_0| ^2+1)
   (1-L)^2\}^{1/2}(|\alpha_0| ^2+1)^{-1}
   (1-L)^{-1}8^{-1/2}.
   \label{IDcoh}
\end{align}

If vacuum state input ($|\alpha_0|=0$) with the loss of $L_1 = L_2 = L$, the phase sensitivity with the ID is found to be%
\begin{align}
\Delta \phi_{\text{L}}^{\text{I}}=&\{\text{csch}^4(2 g)[\csc ^2(\frac{\phi}{2}) (L^2 \cosh (4 g)+2 (1-L) L \cosh (2 g)-(2-L) L)+\sec
   ^2(\frac{\phi}{2}) (L^2 \cosh (4 g)\\\nonumber
   &+(1-L)^2 \cosh (8 g)+2 L (1-L) \cosh (6 g)-1)]+8 (1-L)^2\}^{1/2}
   (1-L)^{-1}8^{-1/2}.
\end{align}

\subsection{Phase sensitivity with the PD with one-coherent-state input with different losses on the two arms}
\label{sec:cohl1l2} If one coherent state input is considered as input, with different losses,
$L_1 \neq L_2$, the phase sensitivity is found to be%
\begin{eqnarray}
\Delta \phi_{\mathrm{L}} = \sqrt{\frac{f_1}{f_2}},
\end{eqnarray}
where
\begin{align}
f_1 = 1 - \frac{4}{c_2}e^{c_1},  \nonumber \\
f_2 = 256 \frac{d_1}{d_2}e^{d_3},
\end{align}
with
\begin{align}
c_1 = &-[2 |\alpha_0|^2 \sinh ^2(2 g) (2 \sqrt{1-L_1} \sqrt{1-L_2} \cos \phi
+L_1+L_2-2)]/[2   \sqrt{1-L_1} \sqrt{1-L_2} \sinh ^2(2 g) \cos \phi +4 L_1
\sinh ^4(g)  \nonumber \\
&+L_2 \sinh ^2(2 g)-\cosh (4 g)-1],  \nonumber \\
c_2 = &(-2 \sqrt{1-L_1} \sqrt{1-L_2} \sinh ^2(2 g) \cos (\phi )-4 L_1 \sinh
^4(g) -L_2 \sinh ^2(2 g)+\cosh (4 g)+1)^2,  \nonumber \\
d_1 = &(1-L_1) (1-L_2) \sinh ^4(2 g) \sin ^2\phi (-4 (|\alpha_0|^2+1) L_1
\cosh (2 g)  +2 \sqrt{1-L_1} \sqrt{1-L_2} \cosh (4 g) \cos \phi -(-L_1-L_2+2)
\cosh (4 g)  \nonumber \\
&-4 |\alpha_0|^2 (1-L_1)-2 \sqrt{1-L_1} \sqrt{1-L_2} \cos \phi
 -3 (1-L_1)-L_2+1)^2,  \nonumber \\
d_2 = &(4 \sqrt{1-L_1} \sqrt{1-L_2} \sinh ^2(2 g) \cos \phi -4 L_1 \cosh (2 g)
-(-L_1-L_2+2) \cosh (4 g)-3 (1-L_1)-L_2+1)^6,  \nonumber \\
d_3 = &[4 |\alpha_0|^2 \sinh ^2(2 g) (-2 \sqrt{1-L_1} \sqrt{1-L_2} \cos \phi
 -L_1-L_2+2)]/[4 \sqrt{1-L_1} \sqrt{1-L_2} \sinh ^2(2 g) \cos \phi
\nonumber \\
& -4 L_1 \cosh (2 g)-(2-L_1-L_2) \cosh (4 g)   -3 (1-L_1)-L_2+1].
\end{align}

\twocolumn



\ifthenelse{\equal{\journalref}{aop}}{%
\section*{Author Biographies}
\begingroup
\setlength\intextsep{0pt}
\begin{minipage}[t][6.3cm][t]{1.0\textwidth} 
  \begin{wrapfigure}{L}{0.25\textwidth}
    \includegraphics[width=0.25\textwidth]{john_smith.eps}
  \end{wrapfigure}
  \noindent
  {\bfseries John Smith} received his BSc (Mathematics) in 2000 from The University of Maryland. His research interests include lasers and optics.
\end{minipage}
\begin{minipage}{1.0\textwidth}
  \begin{wrapfigure}{L}{0.25\textwidth}
    \includegraphics[width=0.25\textwidth]{alice_smith.eps}
  \end{wrapfigure}
  \noindent
  {\bfseries Alice Smith} also received her BSc (Mathematics) in 2000 from The University of Maryland. Her research interests also include lasers and optics.
\end{minipage}
\endgroup
}{}

\end{document}